\documentclass[twocolumn,
aps,
prx,
showpacs,
amsmath,
amssymb,
floatfix,
longbibliography,
eprint]
{revtex4-1}

\usepackage{subfigure}

\usepackage{xcolor}
\definecolor{darkblue}{RGB}{0,0,150}
\definecolor{nightblue}{RGB}{0,0,100}

\usepackage{graphicx,mathtools,bm,bbm}
\usepackage{MnSymbol}
\usepackage[
colorlinks,
citecolor=darkblue,
linkcolor=darkblue,
urlcolor=nightblue]{hyperref}

\usepackage[english]{babel}
\usepackage[babel,kerning=true,spacing=true]{microtype}
\usepackage[utf8]{inputenc}

\usepackage{soul}

\begin{document}
    
\title{Converting electrons into emergent fermions at a \protect\\ superconductor-Kitaev spin liquid interface}
    
\author{Gilad Kishony}
\affiliation{Department of Condensed Matter Physics,
Weizmann Institute of Science,
Rehovot 76100, Israel}
\author{Erez Berg}
\affiliation{Department of Condensed Matter Physics,
Weizmann Institute of Science,
Rehovot 76100, Israel}

\begin{abstract}
We study an interface between a Kitaev spin liquid (KSL) in the chiral phase and a non-chiral superconductor. When the coupling across the interface is sufficiently strong, the interface undergoes a transition into a phase characterized by a condensation of a bound state of a Bogoliubov quasiparticle in the superconductor and an emergent fermionic excitation in the spin liquid. In the condensed phase, electrons in the superconductor can coherently convert into emergent fermions in the spin liquid and vice versa. As a result, the chiral Majorana edge mode of the spin liquid becomes visible in the electronic local density of states at the interface, which can be measured in scanning tunneling spectroscopy experiments. We demonstrate the existence of this phase transition, and the non-local order parameter that characterizes it, using density matrix renormalization group simulations of a KSL strip coupled at its edge to a superconductor. An analogous phase transition can occur in a simpler system composed of a one-dimensional spin chain with a spin-flip $\mathbb{Z}_2$ symmetry coupled to a superconductor.
\end{abstract}

\maketitle

\section{Introduction}

Quantum spin liquids (QSL)~\cite{savary2016quantum,zhou2017quantum,knolle2019field,broholm2020quantum} are fascinating phases of matter where the quantum fluctuations of the spins prevent magnetic ordering even at zero temperature. Instead, these phases are characterized by an underlying topological order in their ground-state wavefunctions, manifested by the presence of fractionalized excitations. 

Among the many possible quantum spin liquids, there is particular interest in gapped phases whose excitations exhibit non-Abelian statistics. A beautiful example of such a phase, proposed by Alexei Kitaev, consists of a spin-$\frac{1}{2}$ degrees of freedom on a honeycomb lattice with strongly anisotropic exchange interactions~\cite{kitaev2006anyons,winter2017models,hermanns2018physics}. In the presence of a magnetic field that breaks time-reversal symmetry, a gap opens in the bulk spectrum, and a non-Abelian Kitaev spin liquid (KSL) phase is formed. The edges of this phase are predicted to host gapless chiral modes of emergent Majorana fermions. Remarkably, the Kitaev Hamiltonian was argued to be approximately realized in certain multi-orbital compounds with strong spin-orbital coupling~\cite{Jackeli_2009}, such as irridates and $\alpha-$RuCl$_3$. In the latter system, a quantized thermal Hall response in the presence of an applied magnetic field, consistent with a chiral Majorana edge mode, has been reported~\cite{kasahara2018majorana}, although these results are still being debated~\cite{Yamashita2020,Bachus2020,Bachus2021}. 

Definitively identifying a quantum spin liquid in experiment is intrinsically hard, due to the lack of magnetic order or any other kind of local order parameter. In addition, the fractionalized excitations of a quantum spin liquid are electrically neutral, and do not couple to conventional experimental probes, making such excitations hard to detect. A variety of experimental signatures of spin liquids have been proposed~\cite{Senthil2001,Norman2009,Mross2011,Potter2013,Huh2013,punk2014topological,Knolle2014,nasu2016fermionic,Werman2018,Gohlke2018,Halasz2019,Pereira2020,Konig2020,Devakul2020}. In particular, it has been pointed out that the edges of a given type of spin liquid can support different kinds of topologically distinct ``boundary phases''~\cite{Kitaev_2003,levin_protected_2013,Barkeshli2013,lichtman2020bulk}. Some of these boundary phases may only be stabilized at the interface between the spin liquid and another phase of matter, such as a magnet or a superconductor~\cite{Barkeshli_2014,aasen_topological_2016}. The properties of such interfaces can provide unique signatures for the presence of a quantum spin liquid in the bulk, as well as clues for its precise nature.

\begin{figure}
\begin{centering}
\includegraphics[trim={5.5cm 6cm 5.5cm 5cm},clip,width=8.7cm,height=5.5cm]{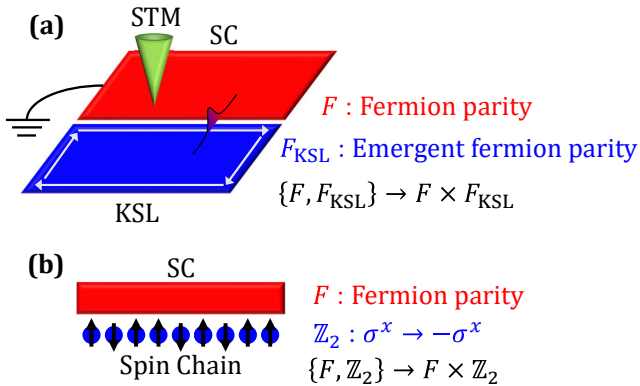}
\par\end{centering}
\caption{The analogy between the hybrid KSL-SC system
(a) and a simple 1d superconductor coupled to a spin
chain (b) presented in terms of their symmetries and
the symmetry broken phase of interest.}
\label{analogy}
\end{figure}

In this study, we propose a method for detection of the characteristic
gapless edge state of the KSL by coupling it to a \emph{topologically trivial}
superconductor (SC) at its
edge. The setup is shown schematically in Fig.~\ref{analogy}a. We show that the KSL-SC interface can
undergo a topological phase transition where the chiral Majorana mode of the KSL becomes hybridized
with the superconducting electrons. In the absence of the coupling to the superconductor, the gapless emergent Majorana fermions at the KSL's edge have no overlap with ordinary electrons, and hence they cannot be detected in a tunneling experiment. In contrast, when the coupling between the KSL and the SC is sufficiently strong, the emergent fermions can coherently convert into electrons~\cite{Barkeshli_2014}. As a result, the gapless edge mode becomes detectable in a scanning tunneling spectroscopy (STM) experiment.

An analogous transition can occur in a simpler system, consisting of a one-dimensional spin chain with a $\mathbb{Z}_2$ spin-flip symmetry coupled to a superconductor (Fig.~\ref{analogy}b). In this system, upon increasing the strength of the interactions between the spin chain and the electrons in the superconductor, a phase transition occurs at which the Jordan-Wigner (JW) fermions of the spin chain become hybridized with electrons in the SC. Formally, this phase transition can be viewed as a symmetry-breaking transition at which both the $\mathbb{Z}_2$ symmetry of the spin chain and the electron number parity symmetry of the superconductor become spontaneously broken, leaving the product of the two symmetries unbroken. Both this transition and the transition at a KSL-SC interface can be described by a non-local order parameter whose expectation value becomes long-ranged in the strongly coupled phase. 

However, unlike in the spin chain case, the transition at a KSL-SC interface does not require any symmetry. Instead, the latter transition can be described in terms of a spontaneous breaking of the \emph{emergent} fermion parity symmetry associated with the fermionic excitations of the KSL and the electron parity symmetry of the superconductor, with their product left unbroken. 

In order to support our conclusions, we study an explicit model of a KSL strip adjacent to a mean-field superconductor (taken to be one-dimensional for simplicity). 
We introduce an interaction between the two systems that couples spins at the edge of the KSL to an electron bilinear operator in the superconductor. 
Solving the model numerically using the density matrix renormalization group (DMRG) technique, we locate the transition as a function of the coupling across the interface. We calculate the non-local order parameter that characterizes the transition, composed of the product of an electron operator in the superconductor and an emergent Majorana fermion operator in the KSL, and show that it becomes long-range ordered in the strong coupling phase.

Our setup is related to that proposed in Ref.~ \cite{Aasen_2020} where an interface between a KSL and a \emph{topological} SC with a counterpropagating chiral edge state was considered. In that system, upon increasing the coupling across the interface past a critical value, the edge modes are gapped out. 
In the present study, the phases on both sides of the transition are gapless (since the interface is chiral). 
Thus, after the transition the gapless edge state of the KSL is not lost, but rather becomes hybridized with the superconducting electrons.

This paper is organized as follows. In Sec. \ref{section:Coupled spin chain and SC} we discuss the one-dimensional toy model in terms of its symmetry-breaking phase transition and its string order parameter. In Sec. \ref{section:Coupled KSL and SC} we present our setup of an interface between a KSL and a SC. We argue for the existence of a phase transition along the edge, and present a non-local order parameter that characterizes it. Finally, Sec. \ref{section:Numerics (DMRG)} describes our numerical DMRG results. In the appendices, we briefly review some properties of the KSL, and analyze the properties of the KSL-SC boundary in an exactly solvable limit. 

\section{Coupled spin chain and SC}
\label{section:Coupled spin chain and SC}

For illustrative purposes, we start by studying a system composed of a spin chain with a $\mathbb{Z}_2$ symmetry, coupled to a superconductor (Fig.~\ref{analogy}b). Despite being much simpler than the KSL edge coupled to a one-dimensional spinless superconductor (Fig.~\ref{analogy}a), which is the main focus of our work, the two problems bear some resemblance to each other in terms of their phase diagram and the interpretation of some of the possible phases. We will comments on the similarities and differences below.  

We consider the following one-dimensional Hamiltonian:
\begin{align}
\label{SC-spin chain}
\mathcal{H}_{1{\rm d}}= & -J\sum_{i}\left(g\sigma_{i}^{z}+\sigma_{i}^{x}\sigma_{i+1}^{x}\right)\nonumber\\
 & -\frac{t}{2} \sum_{i}\left(c_{i}^{\dagger}c_{i+1}^{\vphantom{\dagger}}  + c_{i}^{\dagger}c_{i+1}^{\dagger}+\rm{h.c.}\right) -\mu \sum_i c_{i}^{\dagger}c_{i}^{\vphantom{\dagger}} \nonumber\\
 & +K\sum_{i}\sigma_{i}^{z}\left(c_{i}^{\dagger}c_{i}^{\vphantom{\dagger}} - \frac{1}{2}\right).
\end{align}
where $\sigma^{x,y,z}_i$ are the Pauli matrices acting on a spin-1/2 degree of freedom at a site $i$ in the spin chain, and $c^\dagger_i$ creates an electron at site $i$ in the SC.
The spin chain is governed by a transverse field Ising Hamiltonian with exchange coupling $J$ and a dimensionless coupling constant $g$. $\mu$ is the chemical potential in the superconductor. For simplicity, we take the hopping and the pairing in the superconductor to be both equal to $t$. $K$ denotes the strength of the coupling between the spin chain and the superconductor. Importantly for our discussion, the Hamiltonian \eqref{SC-spin chain} commutes with the fermion parity operator $F$, under which $c_i \rightarrow -c_i$, and with a $\mathbb{Z}_2$ spin flip symmetry $U$ that takes $\sigma_i^x\rightarrow -\sigma_i^x$. 

A convenient way to treat this system is to perform a Jordan-Wigner transformation on the electronic degrees of freedom of the superconductor, mapping the problem onto a system of two coupled spin chains. The transformation is written as
\begin{equation}
    c_{i}=\left(\prod_{j<i}s_j^{z}\right)s_{i}^{-},
\end{equation}
where the fermions are represented by a spin-1/2 degree of freedom with corresponding Pauli matrices $s^{x,y,z}_i$. After the transformation, the Hamiltonian takes the form
\begin{align}
\label{spin chain-spin chain}
\mathcal{H}_{\rm{1d}}	=&-J\sum_{i}\left(g \sigma_{i}^{z}+\sigma_{i}^{x}\sigma_{i+1}^{x}\right)\\&
 +\sum_{i}\left(-\frac{\mu}{2} s_{i}^{z} +t s_{i}^{x}s_{i+1}^{x}\right)-\frac{K}{2}\sum_{i}\sigma_{i}^{z}s_{i}^{z}.\nonumber
\end{align}
This is simply a system of two coupled transverse field Ising models with a $\mathbb{Z}_2\times \mathbb{Z}_2$ symmetry - a system which is intimately related with the Ashkin-Teller model~\cite{Ashkin1943,baxter2016exactly}. The system supports six distinct gapped phases~\cite{verresen2019gapless}. These include a trivial symmetric phase, a symmetry protected topological phase (SPT), three partial symmetry breaking phases, and a fully symmetry-broken phase. Three of the four symmetry broken phases are characterized by having either $\langle \sigma_i^x\rangle \neq 0$, or $\langle s_i^x\rangle \neq 0$, or both. A fourth phase is characterized by $\langle s^x_i \sigma^x_i \rangle \neq 0$, while $\langle \sigma^x_i \rangle = \langle s^x_i \rangle = 0$, such that the two individual $\mathbb{Z}_2$ symmetries are broken, but their product is not. Within the model~\eqref{spin chain-spin chain}, this phase is obtained, e.g., for $g=0$, $\mu=0$, $K\gg J,\,t > 0$. Note that this phase requires a strong coupling between the two spin chains; if $K$ is smaller than $J$ and $t$, each of the two $\mathbb{Z}_2$ symmetries is individually broken. 

The latter phase where the two $\mathbb{Z}_2$ symmetries are broken while their product is preserved is our main focus here; we will show later that an analogous phase can be realized at a boundary between a superconductor and a 2d KSL, even without a global $\mathbb{Z}_2$ symmetry. To understand the properties of this phase in the 1d case, it is useful to note that since the product of the two $\mathbb{Z}_2$ symmetries is unbroken, the phase is also characterized by the following string order parameter:
\begin{equation}
    O_{ij}  = \prod_{i \le l \le j} s^z_l \sigma^z_l,
\end{equation}
such that $\langle O_{ij} \rangle \neq 0$  in the limit $|i-j|\rightarrow \infty$. This is the disorder parameter for the product of the two $\mathbb{Z}_2$ symmetries. Hence, the product of the two order parameters $O_{ij}$ and $\sigma^x_i s^x_i$ is also non-zero. To interpret this order parameter, it is useful to perform a Jordan-Wigner transformation on the original spin chain, mapping it to a system of fermions:
\begin{align}
\sigma_{i}^{z}=&1-2d_{i}^{\dagger}d_{i} \nonumber \\
\sigma_{i}^{x}=&\prod_{j<i}\left(1-2d_{j}^{\dagger}d_{j}\right)\left(d_{i}+d_{i}^{\dagger}\right). 
\end{align}
In terms of the two types of fermions, $c_i$ and $d_i$, the combined order parameter composed of the product of $O_{ij}$ and $\sigma^x_i s^x_i$ is simply the product of the two fermion operators $(d_i + d_i^\dagger) (c_i + c_i^\dagger)$. Therefore, in this phase, the physical electrons in the superconductor hybridize with the JW fermions of the spin chain.

\section{Coupled KSL and SC}
\label{section:Coupled KSL and SC}

We now turn to analyze the interface between a chiral KSL and a superconductor. We show that, analogously to the spin chain-SC system described in the previous section, the electrons in the superconductor can become coherently hybridized with the emergent fermions of the KSL if the coupling between the two subsystems is sufficiently strong. This hybridization onsets at a quantum phase transition along the edge. We identify a non-local string order parameter that becomes long-ranged at the transition. 

\begin{figure}
\begin{centering}
\includegraphics[trim={5.3cm 7.9cm 5cm 7.0cm},clip,width=9cm,height=3.3cm]{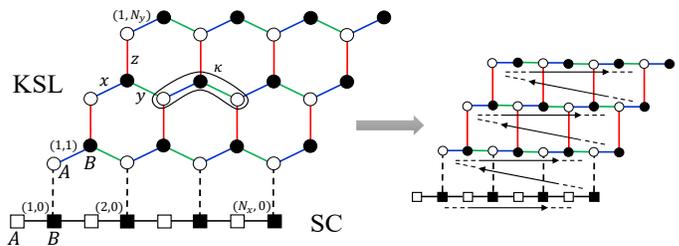}
\par\end{centering}
\caption{\label{full SC KSL system with BC} On the left is a Kitaev spin liquid (circular sites) coupled to a superconductor (square sites). The bonds are colored according to orientation: x-blue, y-green, z-red. The sites are numbered according to their unit cell, $(i,j)$ and their sub-lattice index, $A$ for white and $B$ for black. On the right, the honeycomb lattice is deformed into a brick-wall lattice and the sites are ordered row by row.}
\end{figure}

\subsection{Model for KSL-SC interface}

Our system is shown in Fig.~\ref{full SC KSL system with BC}. A finite strip of KSL in the gapped (chiral) phase is placed near a  superconductor. Since we are interested in phenomena that occur at the interface between the two systems, and for computational simplicity, our model includes only the last row of sites in the superconductor (i.e., we treat the superconductor as being a one-dimensional system with a mean-field pairing potential). Electrons in the  superconductor are coupled to spins in the last row of the KSL strip. The Hamiltonian of the system is written as
\begin{align}
\label{physical operators full Hamiltonian}
\mathcal{H} & = \mathcal{H}_\text{KSL} + \mathcal{H}_\text{SC} + \mathcal{H}_\text{int}.
\end{align}
Here, $\mathcal{H}_{\rm{KSL}}$ is the Hamiltonian of the KSL strip,
\begin{align}
\label{H KSL part}
&\mathcal{H}_\text{KSL} = -J\,\Big[\sum_{\text{\ensuremath{x-}bonds}}\sigma_{(i,j,A)}^{x}\sigma_{(i.j,B)}^{x}
\nonumber
\\
& +\sum_{\text{\ensuremath{y-}bonds}}\sigma_{(i,j,B)}^{y}\sigma_{(i+1,j,A)}^{y}\nonumber+\sum_{\text{\ensuremath{z-}bonds}}\sigma_{(i,j,B)}^{z}\sigma_{(i,j+1,A)}^{z}\,\Big]\nonumber
\\
& -\kappa\sum_{\langle\langle \bm{J},\bm{K},\bm{L}\rangle\rangle}\sigma_{\bm{J}}^x\sigma_{\bm{K}}^y\sigma_{\bm{L}}^z -\sum_{i}h_1^z\sigma_{(i,N_y,B)}^z.
\end{align}
Here, each site of the KSL is labeled by $\bm{I} = (i,j,s)$, where $i \in \{1,\dots,N_x\}$ and $j \in \{1,\dots,N_y\}$ label the unit cell (see Fig.~\ref{full SC KSL system with BC}) and $s=A,B$ labels the two sublattices of the honeycomb lattice. $J$ is the Kitaev anisotropic exchange coupling. $\kappa$ is a three-spin term acting on three neighboring sites denoted by $\langle\langle \bm{J},\bm{K},\bm{L}\rangle \rangle$, arranged as in the configuration shown in Fig.~\ref{full SC KSL system with BC} and every configuration related to that one by a symmetry of the honeycomb lattice. The $\kappa$ term breaks time reversal symmetry, and can be thought of as arising from a magnetic field in the $(1,1,1)$ direction~\cite{kitaev2006anyons}. This term opens a gap in the bulk and drives the system into the chiral phase. We
have also included a Zeeman field $h_1^z$ on the last row of sites in the KSL, in order to prevent a macroscopic accidental degeneracy along that edge. $\mathcal{H}_{\rm{KSL}}$ is exactly solvable by fermionization~\cite{kitaev2006anyons,Chen_2008}. We briefly review the method of solution in Appendix~\ref{app:KSL}. 

The sites of the superconductor are similarly labeled by an index $\bm{I}=(i,0,s)$ with $s=A,B$, i.e., we label the superconducting sites as an additional row below the KSL. Each site has a single (spinless) electronic state, with a corresponding creation operator $c^\dagger_{\bm{I}}$. It is convenient to work with Majorana operators, $\alpha^{\vphantom{\dagger}}_{\bm{I}} = c^\dagger_{\bm{I}} + c^{\vphantom{\dagger}}_{\bm{I}}$, $\beta^{\vphantom{\dagger}}_{\bm{I}} = (c^\dagger_{\bm{I}} - c^{\vphantom{\dagger}}_{\bm{I}})/i$. In terms of these operators, we choose the Hamiltonian of the superconductor to be of the form:
\begin{align}
\label{H SC part}
\mathcal{H}_\text{SC} & = 
 -i\gamma\sum_{i}\alpha_{(i,0,A)}\alpha_{(i,0,B)}\nonumber
 \\
 &  +i\gamma\sum_{i}\alpha_{(i,0,B)}\alpha_{(i+1,0,A)}\nonumber
 \\
 & -i\varepsilon_A\sum_{i}\beta_{(i,0,A)}\alpha_{(i,0,A)}+i\varepsilon_B\sum_{i}\beta_{(i,0,B)}\alpha_{(i,0,B)}.
\end{align}
which corresponds to a combination of hopping terms and pairing potentials in terms of the electronic operators $c_{\bm{I}}$. The reason for this choice will become apparent later. 

The interaction between the superconductor and the KSL is written as
\begin{align}
\label{H int part}
\mathcal{H}_\text{int} & = iK\sum_{i}\sigma_{(i,1,A)}^{z}\beta_{(i,0,B)}\alpha_{(i,0,B)},
\end{align}
which corresponds to a coupling of the spins in the first row of the KSL to the density of electrons in the superconductor, $i\alpha_{\bm{I}}\beta_{\bm{I}} = 2c_{\bm{I}}^\dagger c_{\bm{I}}-1$. Note that this term is not forbidden, since time reversal symmetry is broken in our system. Of course, the fermion parity of electrons in the superconductor is conserved. In additional, the Hamiltonian is invariant under a spin rotation by $\pi$ around the $z$ axis.

\subsection{Phase transition along the interface and string order parameter}
In the absence of coupling between the KSL edge and the SC ($K=0$), the KSL edge supports a gapless chiral mode, while the SC is gapped. The electrons in the superconductor are expected to remain gapped for non-zero $K$, as long as $K$ is sufficiently small. However, beyond a certain value of $K$, a phase transition may occur along the interface, beyond which the electrons in the superconductor become hybridized with the emergent fermions of the KSL. We now identify a non-local string order parameter that characterizes the phase transition, and argue that such a transition must occur in the model (\ref{physical operators full Hamiltonian}). In Sec.~\ref{section:Numerics (DMRG)} we will study the phase transition numerically, using DMRG.

The order parameter that we expect to become long-range-ordered at large $K$ is given by the product of the electron operator in the superconductor times a Majorana operator of an emergent fermions of the KSL near the boundary with the SC. 
For example, we may use the following operator:
\begin{align}
O_0(i) = i\alpha_{(i,1,A)}\alpha_{(i,0,B)},
\end{align}
where $\alpha_{(i,0,B)}$ is a Majorana operator in sublattice $B$ of the superconductor, and $\alpha_{(i,1,A)}$ is a Majorana operator defined through a Jordan-Wigner transformation on the $j=1$ row of spins in the KSL (see Appendix~\ref{app:KSL} for an explicit definition). 
In the large $K$ phase, we expect the correlation function of $O_0$ to become long-range ordered along the boundary, i.e. $S_0(i,i')\equiv \langle O_0(i) O_0(i') \rangle$ should approach a non-zero constant at large $|i-i'|$. 
In terms of the spin operators of the KSL, $S_0(i,i')$ is a non-local string order parameter. Explicitly,
\begin{align}
S_0(i,i') &=   \langle \sigma_{(i,1,A)}^{x}\sigma_{(i,1,B)}^{z} \prod_{i<k<i'}\left(\sigma_{(k,1,A)}^{z}\sigma_{(k,1,B)}^{z}\right)\sigma_{(i',1,A)}^{y}\nonumber\\
&\times i\alpha_{(i,0,B)}\alpha_{(i',0,B)}\rangle.
\label{eq:SOP}
\end{align}
In addition, we define a string order parameter in terms of a product of an electronic operator in the SC times a Majorana fermion operator at depth $D$ in the KSL: $O_D(i) = i\alpha_{(i-D,2D+1,A)}\alpha_{(i,0,B)}$. The correlation function of this operator should also become long-ranged in the large $K$ phase. However, as we shall demonstrate below, its asymptotic magnitude at large $|i-i'|$ decays exponentially with $D$.

Clearly, when $K=0$, $S_0(i,i')$ decays exponentially at large distances, since $S_0$ decomposes into a product of a correlator in the KSL times the Greens' function of the SC, $\langle i\alpha_{(i,0,B)} \alpha_{(i',0,B)}\rangle$, and the latter correlation function decays exponentially since the fermions in the SC are gapped. Conversely, in the opposite limit of large $K$, we now argue that $S_0$ becomes long-ranged. 
This can be seen by first setting $\varepsilon_B=0$. The Hamiltonian (\ref{physical operators full Hamiltonian}) is then exactly solvable even for $K\ne 0$, as we show in Appendix~\ref{app:solvable}. This is because for $\varepsilon_B=0$, upon fermionization of the KSL spins, the degrees of freedom of the superconductor can be regarded as an additional row of the KSL. In this case, the correlation function $S_0$ can be shown to be long-ranged. Introducing a non-zero $\varepsilon_B$ diminishes the magnitude of the correlation function, but does not decrease it immediately to zero (see Appendix \ref{app:solvable}). 

Hence, for $\varepsilon_B \ne 0$, there must be a phase transition at an intermediate value of $K$ where $S_0$ becomes non-zero. The properties of this transition are studied using DMRG in the next section.

\section{DMRG simulations}
\label{section:Numerics (DMRG)}


In order to demonstrate the existence of a phase transition on the KSL-SC boundary, we performed DMRG simulations of the model (\ref{physical operators full Hamiltonian}) on a finite strip. Measuring the non-local order parameter at different values of the coupling $K$ across the interface, we identify a transition beyond which the order parameter becomes long-ranged.  The DMRG calculations were performed using the TeNPy Library \cite{tenpy}.


In all the following calculations we fix the parameters $J=1$, $\gamma=1$, $\kappa=0.1$, $h_1^z=\varepsilon_A=0.2$, $\varepsilon_B=0.05$ and we vary $K$ and the system size $N_x \times N_y$. We order the sites in the matrix product state column by column: $(0,0,A), (0,1,A),\dots, (0,N_y,A), (0,0,B),(0,1,B)\dots$ The bond dimension was increased every few sweeps and needed to reach a value of up to 1000 for convergence. The error bars drawn for the results are calculated as a difference between the value reached at the final bond dimension and that reached at the second to last value. 

\begin{figure}
\begin{centering}
\includegraphics[width=8.7cm,height=5.3cm]{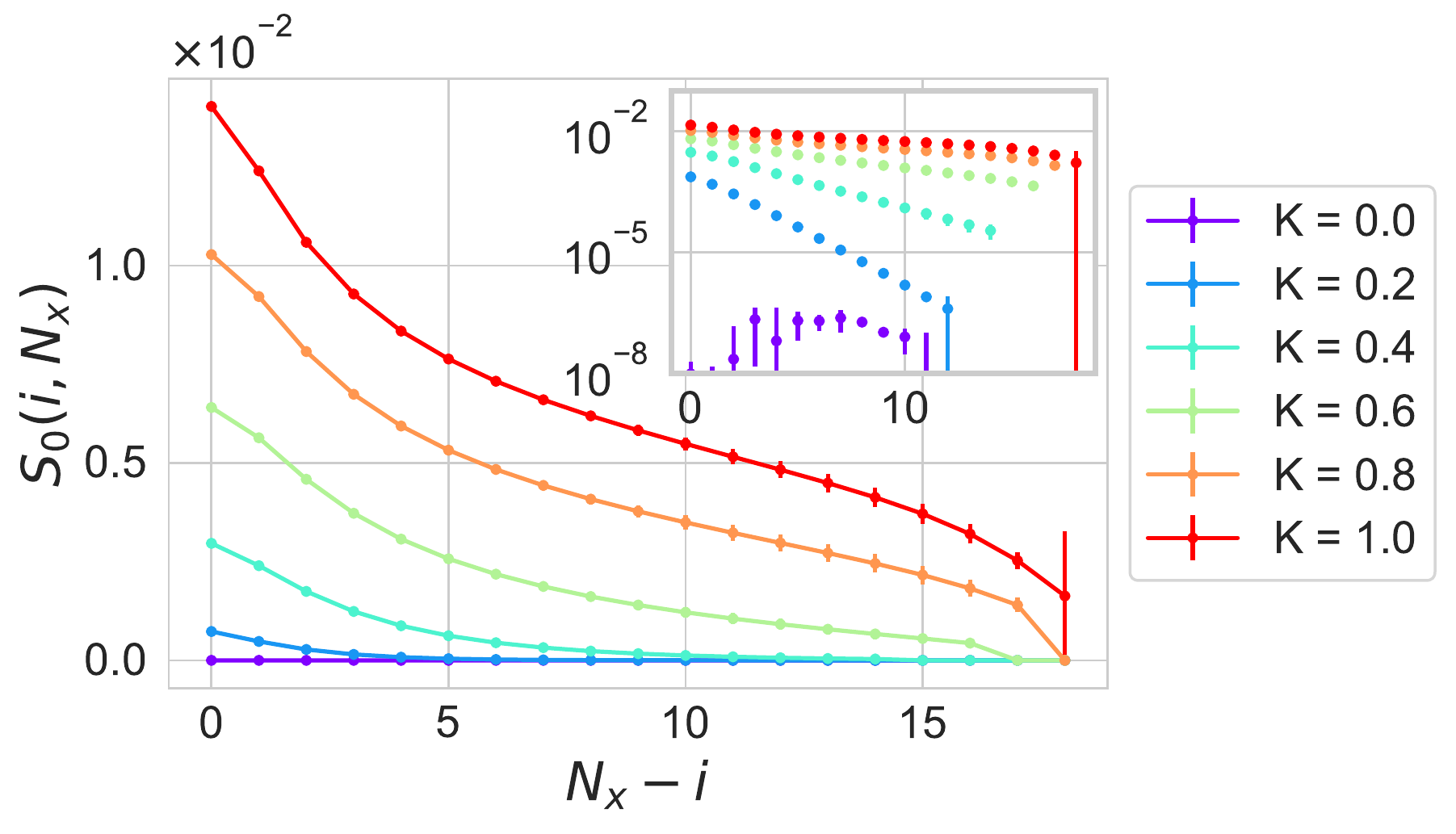}
\par\end{centering}
\caption{\label{SOP vs x lin and log}The correlation of the string order parameter of the KSL fermion times the SC fermion as a function of distance for different values of  KSL-SC coupling in a system of size $N_x=20,N_y=2$ at depth $0$. The inset shows the same in linear-log scale.}
\end{figure}

Fig. \ref{SOP vs x lin and log} shows the string correlation function $S_0(i,N_x)$ [Eq. (\ref{eq:SOP})] as a function of distance $N_x-i$ in a system with $N_x=20$, $N_y=2$, for different values of KSL-SC coupling $K$. The inset shows the same data in a linear-log scale. It is clearly evident that in the limit of small $K$, the string order parameter is exponentially decaying with distance, while beyond $K\approx 0.4$ the correlation length becomes comparable to system size. 

\begin{figure}
\begin{centering}
\includegraphics[trim={10.2cm 4.0cm 9.1cm 4.7cm},clip,width=8.5cm,height=12.5cm]{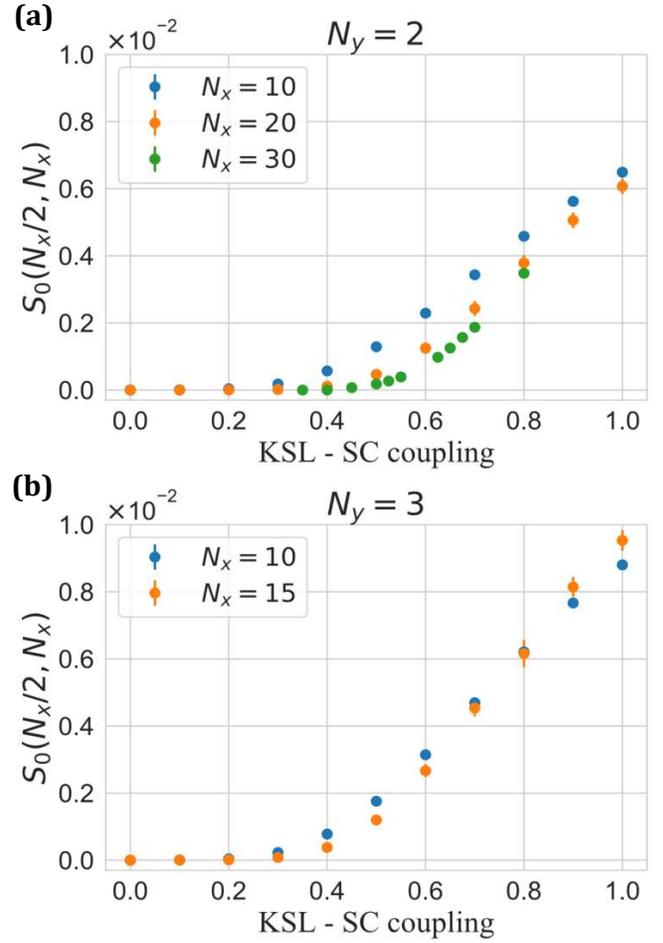}
\par\end{centering}
\caption{\label{SOP vs. K}The correlation of string order parameter of the KSL fermion times the SC fermion at half the system length as a function of the KSL-SC coupling for different system lengths. The results are shown for a system of width $N_y=2$ (a), and for $N_y=3$ (b).}
\end{figure}

To support the existence of a phase with long-range string order, we investigate the dependence of the string correlation function at half the system, $S_0(N_x/2,N_x)$, on $K$ for different system sizes.
This is shown in Fig. \ref{SOP vs. K}(a,b) for $N_y=2$ and $N_y=3$, respectively. For both values of $N_y$, the behavior is consistent with a continuous transition at $K_c\approx 0.4$ where the string correlation function becomes non-zero in the limit of large $N_x$.  



\begin{figure}
\begin{centering}
\includegraphics[width=8.7cm,height=5.3cm]{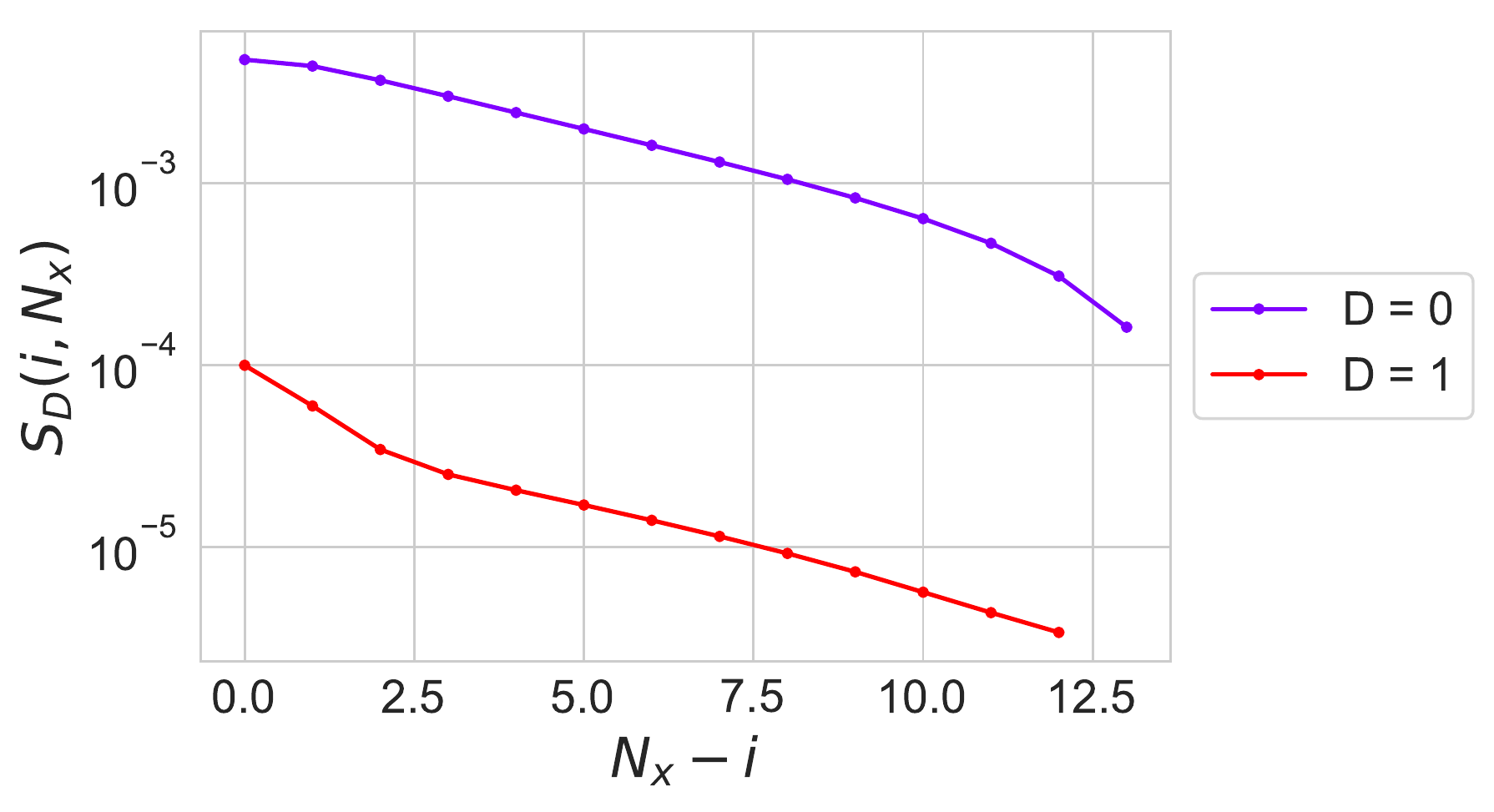}\\
\par\end{centering}
\caption{\label{SOP vs. K depth 0 and 2} The correlation of the string order parameter of the KSL fermion at depths $0,1$ times the SC fermion as a function of distance for $K=0.5$ in a system of size $N_x=15, N_y=3$.}
\end{figure}

Next, we investigate the dependence of the string order parameter on the depth $D$ at which the KSL fermion is taken. Figure \ref{SOP vs. K depth 0 and 2} compares the values of the string order parameter at depths $D=0$ and $1$. The order parameter decays rapidly with depth, as expected due to the strong localization of the KSL edge state.

\begin{figure}
\begin{centering}
\includegraphics[trim={10.2cm 5.4cm 9cm 4.9cm},clip,width=8.5cm,height=12.2cm]{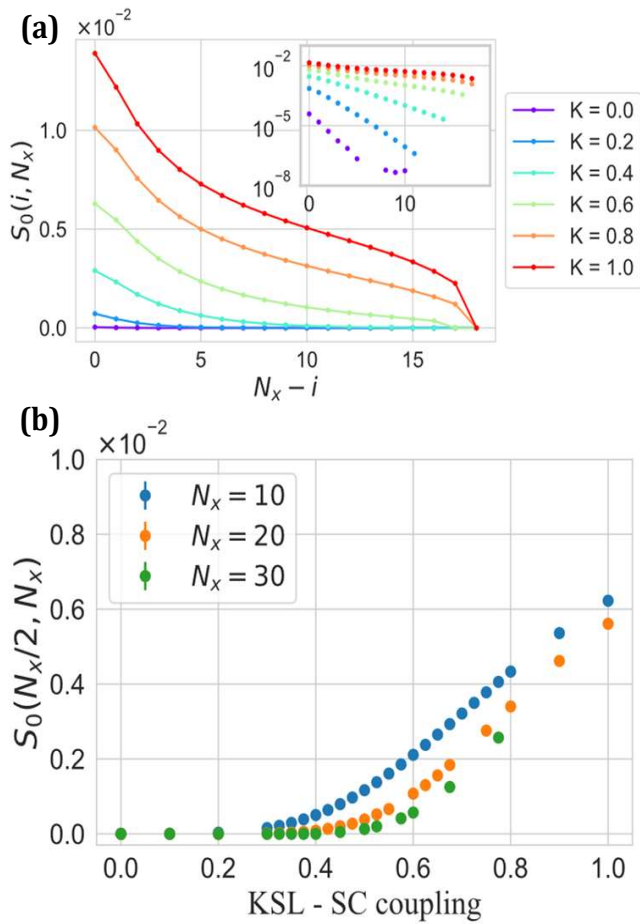}
\par\end{centering}
\caption{\label{with Kx}Breaking the $\mathbb{Z}_2$ symmetry in the KSL with $K' = 0.05$ leaves the phase transition intact. The correlation of the string order parameter of the KSL fermion times the SC fermion as a function of distance for different values of  KSL-SC coupling in a system of size $N_x=20,N_y=2$ at depth $0$ is shown in (a). The inset shows the same in linear-log scale. The correlation of string order parameter of the KSL fermion times the SC fermion at half the system length as a function of the KSL-SC coupling for different system lengths with $N_y=2$ is presented in (b).}
\end{figure}

As mentioned above, the model (\ref{physical operators full Hamiltonian}) possesses a $\mathbb{Z}_2$ symmetry associated with a $\pi$ rotation around the $z$ axis in spin space. However, unlike in the one-dimensional model described in Sec.~\ref{section:Coupled spin chain and SC}, we do not expect this symmetry to play a crucial role in the transition along the KSL-SC interface. In particular, the transition should exist even if the symmetry is broken. To demonstrate this, we added to the Hamiltonian (\ref{physical operators full Hamiltonian}) a term $\mathcal{H}'_\text{int} = iK'\sum_{i}\sigma_{(i,1,A)}^{x}\beta_{(i,0,B)}\alpha_{(i,0,B)}$ with $K' = 0.05$. The results for the string correlation function in this case are shown in Fig.~\ref{with Kx}. As can be seen in the figure, there still is a clear transition near $K=0.4$ where $S_0(i,N_x)$ becomes long-ranged.

\begin{figure}
\begin{centering}
\includegraphics[width=8.7cm,height=5.3cm]{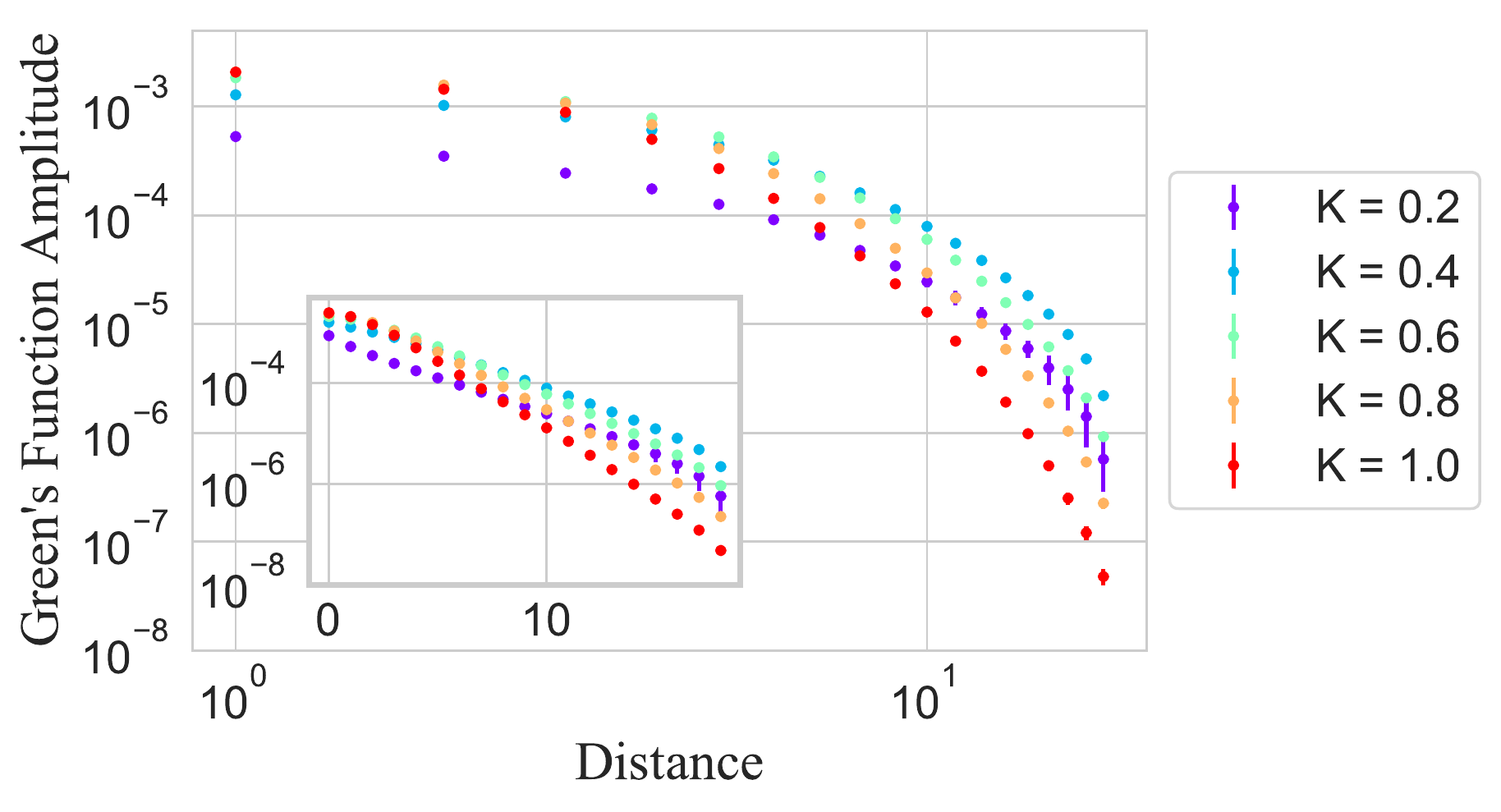}
\par\end{centering}
\caption{\label{green's function dmrg}The Green's function the $\beta_{(i,0,A)}$ electron in the SC as a function of distance for different values of KSL-SC coupling in a system of size $N_x=20,N_y=2$ in log-log scale. The inset shows the same in linear-log scale.}
\end{figure}

Finally, in the phase where the string correlation function becomes long-ranged, we expect the electronic Green's function to display power law correlations along the interface. This is because the electronic quasiparticles hybridize with the emergent gapless Majorana fermions at the edge of the KSL. In our simulations, the electronic Green's function is found to decay rapidly along the interface even for $K>K_c$, and no power law could be discerned (see Fig.~\ref{green's function dmrg}). Examining the electronic Green's function of the
$\beta_{(i,0,A)}$ electron in the SC in the exactly solvable limit (Appendix~\ref{app:solvable}), ${\varepsilon}_B=0$, we find that a width of at least $N_y \gtrsim 5$ is needed before the power law decay along the interface can be clearly seen, explaining this apparent discrepancy.

\section{Summary}

To summarize, in this work, we have shown that the boundary between a superconductor and a chiral Kitaev spin liquid displays a phase transition as a function of the coupling between the two systems. In the strong coupling phase, the electrons in the superconductor become hybridized with the emergent fermions of the spin liquid. Hence, in this phase, electrons can tunnel into the chiral gapless edge modes of the spin liquid. The resulting gapless electronic density of states at the interface should be detectable in STM experiments. 

A similar transition is possible at the interface of a superconductor with other spin liquid phases, such as a gapless (Dirac) Kitaev spin liquid. More generally, any spin liquid that has emergent fermionic bulk excitations can support this type of interface. 

Our numerical results suggest that the transition at the interface of a superconductor with a chiral KSL is continuous. This is an unusual one-dimensional quantum critical point, since it separates two chiral phases. Understanding the universality class of this transition is an interesting direction for future investigation. 

\acknowledgements

We thank Maissam Barkeshli, Steve Kivelson, and Yuji Matsuda for stimulating discussions. This work was supported by the European Research Council (ERC) under grant HQMAT (Grant Agreement No. 817799), the Israel Science Foundation Quantum Science and Technology initiative (grant no. 2074/19), CRC 183 of the Deutsche Forschungsgemeinschaft (Project C02) and a Research grant from Irving and Cherna Moskowitz.

\appendix

\section{Kitaev honeycomb model}
\label{app:KSL}
The Kitaev honeycomb model consists of spin-halves, located at the vertices of a
honeycomb lattice. The bonds in this lattice are divided into three
distinct sets according to their orientation and are called ``x-bonds'', ``y-bonds''
and ``z-bonds''. The Hamiltonian is given as follows:
\begin{align}
\label{basic KSL Hamiltonian}
\mathcal{H} =& -J_{x}\sum_{\text{\ensuremath{x-}bonds}}\sigma_{(i,j,A)}^{x}\sigma_{(i,j,B)}^{x}\nonumber
\\
& -J_{y}\sum_{\text{\ensuremath{y-}bonds}}\sigma_{(i,j,B)}^{y}\sigma_{(i+1,j,A)}^{y}\nonumber
\\
& -J_{z}\sum_{\text{\ensuremath{z-}bonds}}\sigma_{(i,j,B)}^{z}\sigma_{(i,j+1,A)}^{z}\nonumber
\\
& 
-\sum_{\bm{I}}\left(h_{x}\sigma_{\bm{I}}^{x}+h_{y}\sigma_{\bm{I}}^{y}+h_{z}\sigma_{\bm{I}}^{z}\right),
\end{align}
where $J_x,J_y,J_z$ are orientation-dependent coupling strengths and
$\boldsymbol{h}$ is a Zeeman field strength. The strong frustration
arising from these competing couplings suppresses the tendency for
conventional symmetry-breaking order.

With only the $J$ terms present, the ground state and all the excited
states of this system are exactly known.

\subsection{Solution by fermionization}

Before introducing our full model for a coupled KSL and SC, we briefly provide the solution of the exactly solvable KSL model by fermionization using the method described in Ref.~\cite{Chen_2008}. This method will be convenient for expressing the couplings of our interacting model and will also provide us with a useful exactly solvable limit.

We perform a Jordan-Wigner transformation defined by deforming the lattice into a ``brick-wall" geometry, as shown in Fig.~\ref{full SC KSL system with BC}. We order the sites in a 1d contour row by row (see figure), and replace the spin operators with complex fermions $c,c^{\dagger}$, defined as
\begin{align}
&\sigma_{(i,j,s)}^{x}+i\sigma_{(i,j,s)}^{y}=\nonumber\\
  &2\left(\prod_{j'<j,i',s'}\sigma_{(i',j',s')}^{z}\right)\left(\prod_{i'<i,s'}\sigma_{(i',j,s')}^{z}\right)\left(\prod_{s'<s}\sigma_{(i,j,s')}^{z}\right)c_{(i,j,s)}^{\dagger},
\end{align}
\begin{align}
\sigma_{\bm{I}}^{z}=2c_{\bm{I}}^{\dagger}c_{\bm{I}}-1.
\end{align}
With this definition, and setting $\bm{h}=0$, the Hamiltonian in Eq. \eqref{basic KSL Hamiltonian} becomes
\begin{align}
\mathcal{H} & =J_{x}\sum_{\text{\ensuremath{x-}bonds}}\left(c^{\dagger}-c\right)_{(i,j,A)}\left(c^{\dagger}+c\right)_{(i,j,B)}\nonumber\\
 & -J_{y}\sum_{\text{\ensuremath{y-}bonds}}\left(c^{\dagger}+c\right)_{(i,j,B)}\left(c^{\dagger}-c\right)_{(i+1,j,A)}\nonumber\\
 & -J_{z}\sum_{\text{\ensuremath{z-}bonds}}\left(2c^{\dagger}c-1\right)_{(i,j,B)}\left(2c^{\dagger}c-1\right)_{(i,j+1,A)}
\end{align}

Next, we introduce the Majorana fermions $\alpha$ and $\beta$ for sites in the $A$ sub-lattice:
\begin{align}
\alpha_{(i,j,A)}=\frac{1}{i}\left(c-c^{\dagger}\right)_{(i,j,A)},\text{    }&\beta_{(i,j,A)}=\left(c+c^{\dagger}\right)_{(i,j,A)},
\end{align}
and for sites in the $B$ sub-lattice
\begin{align}
\beta_{(i,j,B)}=\frac{1}{i}\left(c-c^{\dagger}\right)_{(i,j,B)},\text{ }&\alpha_{(i,j,B)}=\left(c+c^{\dagger}\right)_{(i,j,B)}
\end{align}

In these terms, the Hamiltonian becomes:
\begin{align}
\label{KSL Hamiltonian in terms of alpha and beta}
\mathcal{H} & =-iJ_{x}\sum_{\text{\ensuremath{x-}bonds}}\alpha_{(i,j,A)}\alpha_{(i,j,B)}\nonumber\\
 & +iJ_{y}\sum_{\text{\ensuremath{y-}bonds}}\alpha_{(i,j,B)}\alpha_{(i+1,j,A)}\nonumber\\
 & -J_{z}\sum_{\text{\ensuremath{z-}bonds}}\left(\beta\alpha\right)_{(i,j,B)}\left(\beta\alpha\right)_{(i,j+1,A)}.
\end{align}

It is now clear that the set of operators
\begin{align}
u_{i,j}=i\beta_{(i,j,B)}\beta_{(i,j+1,A)},
\end{align}
defined on all $z$-bonds labeled by the coordinate of the site in the $B$ sublattice $i,j$, commute with the Hamiltonian and with each other. Replacing these operators with their eigenvalues $u_{i,j}=\pm1$ leaves us with the following easily solvable quadratic Hamiltonian:
\begin{align}
\mathcal{H} & _{u}=-iJ_{x}\sum_{\text{\ensuremath{x-}bonds}}\alpha_{(i,j,A)}\alpha_{(i,j,B)}\nonumber\\
 & +iJ_{y}\sum_{\text{\ensuremath{y-}bonds}}\alpha_{(i,j,B)}\alpha_{(i+1,j,A)}\nonumber\\
 & -i\sum_{\text{\ensuremath{z-}bonds}}J_{z}u_{i,j}\alpha_{(i,j,B)}\alpha_{(i,j+1,A)}
\end{align}

The ground state is found in the vortex-free sector given by $u_{i,j}=1$ as follows from a theorem proved by Lieb \cite{Lieb_1994}.

\subsection{Zeeman field perturbation}

 Although the model is not exactly solvable in the presence of the Zeeman field, it can be solved by perturbation theory where the relevant leading order correction to the low-energy effective Hamiltonian is given by the following three-spin term~\cite{kitaev2006anyons}:
\begin{equation}
\label{three spin term}
\mathcal{H}_{eff}^{(3)}\sim-\kappa\sum_{(\bm{J},\bm{K},\bm{L})\in\Omega}\sigma_{\bm{J}}^x\sigma_{\bm{K}}^y\sigma_{\bm{L}}^z
\end{equation}
where $\kappa=h_xh_yh_z/J^2$, and $\Omega$ is the set of triplets of sites equivalent by symmetry to $\bm{J}=(1,1,A),\bm{K}=(1,1,B),\bm{L}=(1,2,A)$. This three-spin term can be written in terms of the fermionic operators as
\begin{equation}
\label{three spin term for fermions}
\mathcal{H}_{eff}^{(3)}\sim-\kappa\sum_{(\bm{J},\bm{K},\bm{L})\in\Omega}\alpha_{\bm{J}}\beta_{\bm{K}}\alpha_{\bm{L}}\beta_{\bm{L}}.
\end{equation}
Upon restricting to the ground state sector, this simply becomes a second-nearest-neighbor hopping of the $\alpha$ fermions,
\begin{equation}
\label{three spin term for fermions}
\mathcal{H}_{eff}^{(3)}\sim\kappa\sum_{(\bm{J},\bm{K},\bm{L})\in\Omega}i\alpha_{\bm{J}}\alpha_{\bm{L}}.
\end{equation}

\section{KSL-SC model in the exactly solvable limit and edge states}
\label{app:solvable}
In this Section we examine the coupled KSL-SC model given by Eq. \eqref{physical operators full Hamiltonian} in the exactly solvable limit of $\varepsilon_B\rightarrow0$. We extract the wavefunctions and the dispersion of the edge
states and the electronic Green's function. We solve the model with periodic boundary conditions lengthwise: $(i,j,s)=(i+N_x,j,s)$. 
These results are useful in establishing the existence of the strongly coupled KSL-SC interface, and in assessing the role of finite size effects on this phase in a narrow strip geometry. 

The model is solved by writing the fermionized Hamiltonian in the ground state sector (without fluxes) in terms of operators Fourier transformed in the $x$ direction.
\begin{align}
\mathcal{H} & =\frac{1}{2}i\sum_{\ell_1,\ell_2}\sum_{q_{x}}\tilde{A}_{\ell_{1},\ell_{2}}(q_{x})a_{q_{x},\ell_{1}}^{\dagger}a_{q_{x},\ell_{2}},
\end{align}
where
\begin{equation}
a_{q_{x},\ell}=\frac{1}{\sqrt{2N_{x}}}\sum_{i_x}e^{-iq_{x}\cdot i_x} a_{i_x,\ell},
\end{equation}
and we have defined $a_{q_{x},\ell}$ as
\begin{align}
&a_{i_x,\ell=\{1,\dots,2N_y+4\}}\equiv\Big(\beta_{(i_x,0,A)},\alpha_{(i_x,0,A)},\alpha_{(i_x,0,B)},\nonumber\\
&\alpha_{(i_x,1,A)},\alpha_{(i_x,1,B)},\dots,\alpha_{(i_x,N_y,A)},\alpha_{(i_x,N_y,B)},\beta_{(i_x,N_y,B)}\Big).
\end{align}
The matrix $\tilde{A}(q_x)$ is given by
\begin{align}
i\tilde{A}(q_{x})=\begin{pmatrix}0 & i\chi & 0\\
-i\chi & \lambda & is & 0\\
0 & -is & -\lambda & ir & 0\\
 & 0 & -ir & \lambda & is & -\zeta\\
 &  & 0 & -is & -\lambda & ir & \zeta\\
 &  &  & -\zeta & -ir & \lambda & is & -\zeta\\
 &  &  &  & \zeta & -is & -\lambda &  & \ddots\\
 &  &  &  &  & -\zeta &  & \ddots &  & 0\\
 &  &  &  &  &  & \ddots &  &  & -i\chi\\
 &  &  &  &  &  &  & 0 & i\chi & 0
\end{pmatrix},
\end{align}
where $r=2J$, $s=-4J\cos(\frac{q_x}{2})$, $\lambda=4\kappa\sin(q_x)$, $\zeta=4\kappa\sin(\frac{q_x}{2})$ and $\chi=-2h_1^z=-2\varepsilon_A$ assuming $J_x=J_y=J_z=J$. 

Next we diagonalize $i\tilde{A}(q_{x})$ as
\begin{align}
i\tilde{A}(q_{x})=U(q_{x})D(q_{x})U^{\dagger}(q_{x})
\end{align}
where $D(q_{x})$ is real and diagonal, and $U$ is unitary. Substituting this leads to
\begin{align}
\mathcal{H}=\frac{1}{2}\sum_{m}\sum_{q_{x}}b_{q_{x},m}^{\dagger}D_{m,m}(q_{x})b_{q_{x},m}
\end{align}
where
\begin{align}
b_{q_{x},m}=\sum_{\ell}U^{\dagger}_{m, \ell}(q_{x})\,a_{q_{x},\ell}
\end{align}

The eigenvectors given as columns of $U$ correspond to the amplitudes of the different eigenstates of the system on each row of sites, and the diagonal of $D(q_{x})$ gives their dispersions.

Next, we calculate the Green's function of the $\beta_{(i,0,A)}$ electron in the SC by expressing the original fermions in terms of the diagonalized ones. Since we are interested in the ground state, we have
\begin{align}
\langle b_{q_{x,1},m}^{\dagger}b_{q_{x,2},n}\rangle=\Theta[-D_{m,m}(q_{x1})]\delta_{mn}\delta_{q_{x,1}q_{x,2}}.
\end{align}
Therefore, the Green's function is given by
\begin{align}
\label{eq:GF}
\begin{gathered}
\langle \beta_{(i_{1},0,A)}\beta_{(i_{2},0,A)}\rangle= \\
\frac{2}{N_{x}}\sum_{q_{x}}\sum_{m}e^{iq_{x}\cdot\left(i_{2}-i_{1}\right)}U^{\vphantom{\dagger}}_{1,m}(q_x)U_{m,1}^{\dagger}(q_x)\Theta[-D_{m,m}(q_{x})]
\end{gathered}
\end{align}

\begin{figure*}[t]
  \begin{centering}
  \subfigure{\includegraphics[scale=0.39]{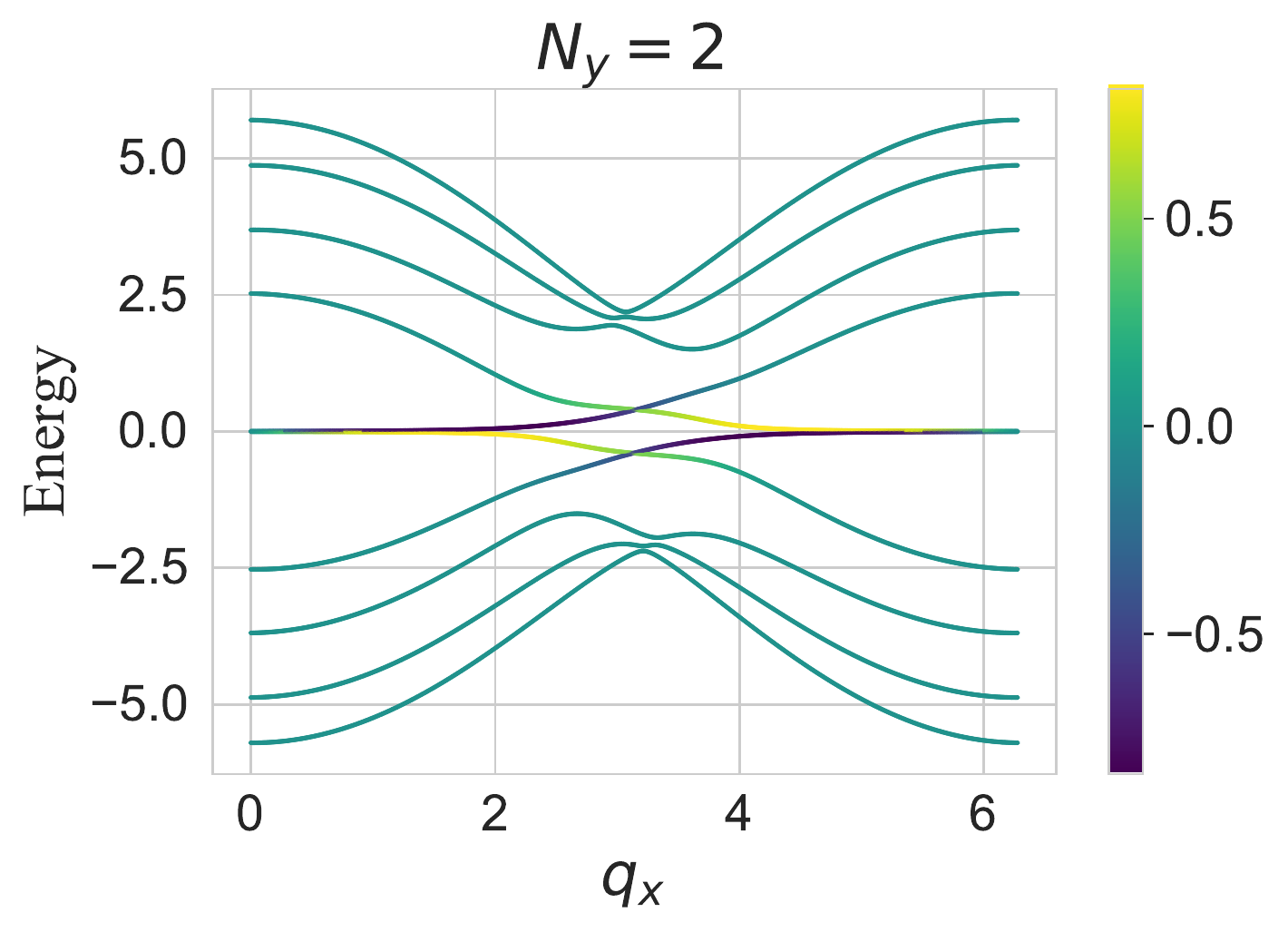}}\quad
  \subfigure{\includegraphics[scale=0.39]{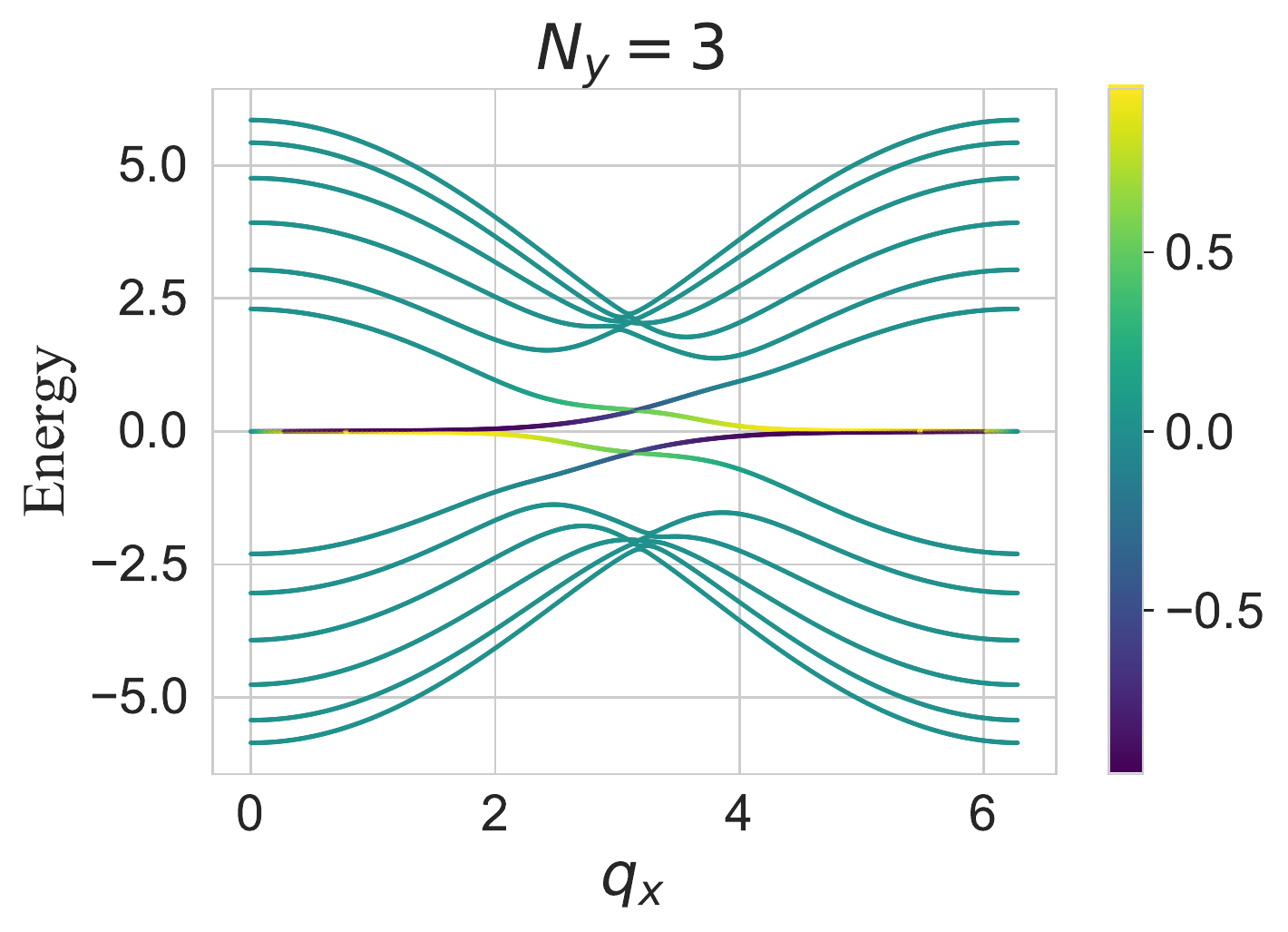}}\quad
  \subfigure{\includegraphics[scale=0.39]{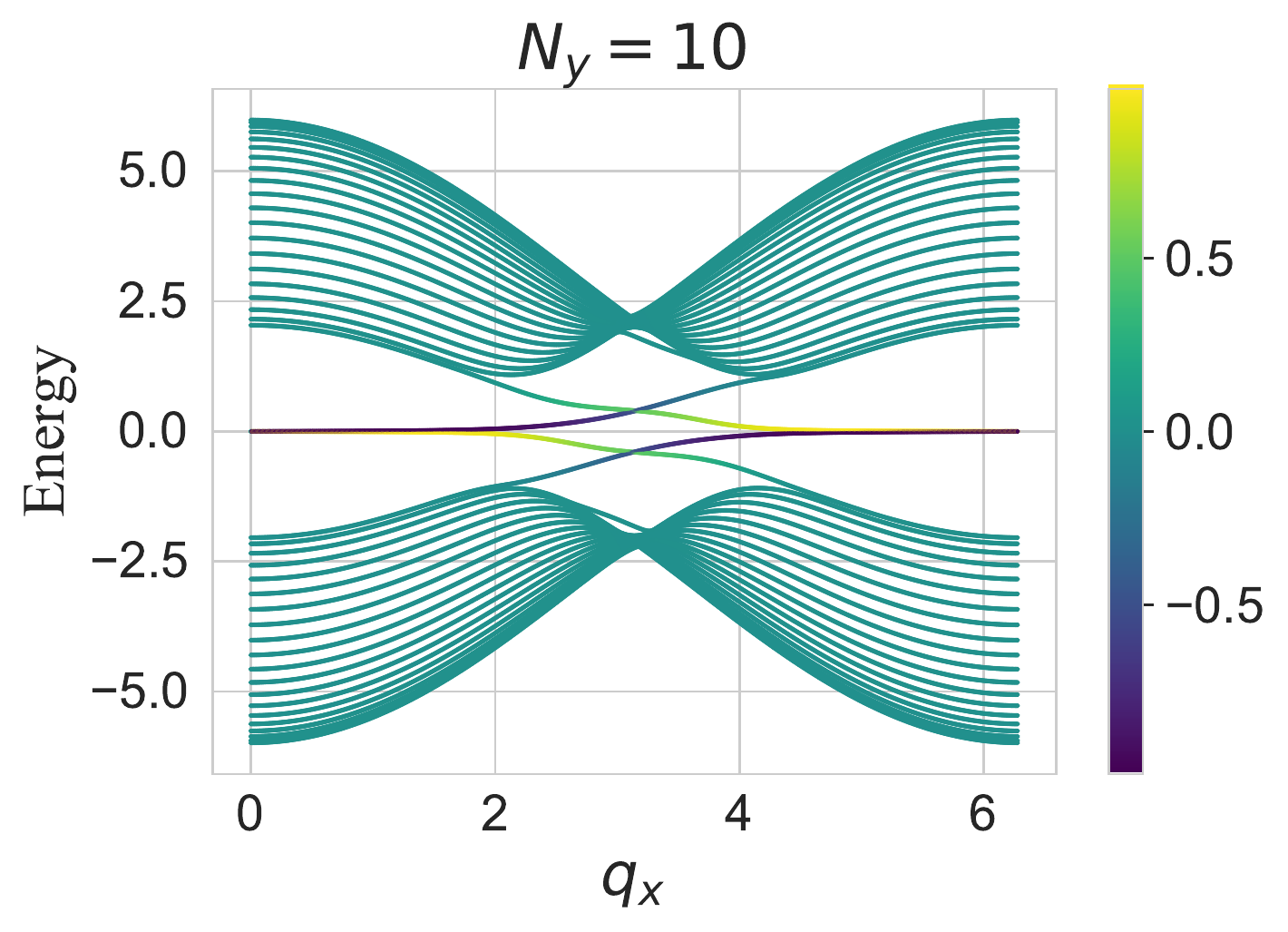}}
  \par\end{centering}
  \caption{\label{dispersion in exactly solvable limit} Dispersion of the fermionic states in a KSL-SC strip in the exactly solvable limit ($\varepsilon_B=0$). On the left the system width is $N_y=2$, in the center $N_y=3$, and on the right $N_y=10$. Points are colored according to the degree of confinement of the wavefunctions to the edge (given by the weight on the top row of $\beta$ operators minus the weight on the bottom row of $\beta$ operators), such that yellow curves correspond to states localized at the top, and blue corresponds to states at the bottom.}
\end{figure*}

We diagonalize the Hamiltonian numerically for a system
of length $N_{x}=1000$ and various widths $N_{y}$. All the other
parameters are set to the same values used in the DMRG simulations, specifically with $K=J_z=1$. In Fig.~\ref{dispersion in exactly solvable limit} we present the dispersion in systems of width $N_y=2,3,10$. The overlap of the eigenstates with the $\beta$ operators on the edges is depicted by the coloring of the curves.

Even in very narrow systems ($N_y=2,3$), the chiral edge states can clearly be
seen both in terms of their dispersion and their localization to the edge. Where
there are avoided crossings between the edge states at the top and
bottom of the strip they become hybridized, but their weight remains
on the edges. The
region in momentum space where the edge states are significantly hybridized becomes smaller as $N_y$ increases,
since the matrix element between the edge states at the opposite edges decays with $N_y$.

Decreasing the three-spin term $\kappa$ results in a smaller gap opening in the bulk which leads to a smaller dispersion of the edge states
and more hybridization between the two edges. 
With our choice of parameters,
the dispersion of the edge states is quite flat near $q_{x}=0$. 

\begin{figure}
\begin{centering}
\includegraphics[width=8.5cm,height=5.8cm]{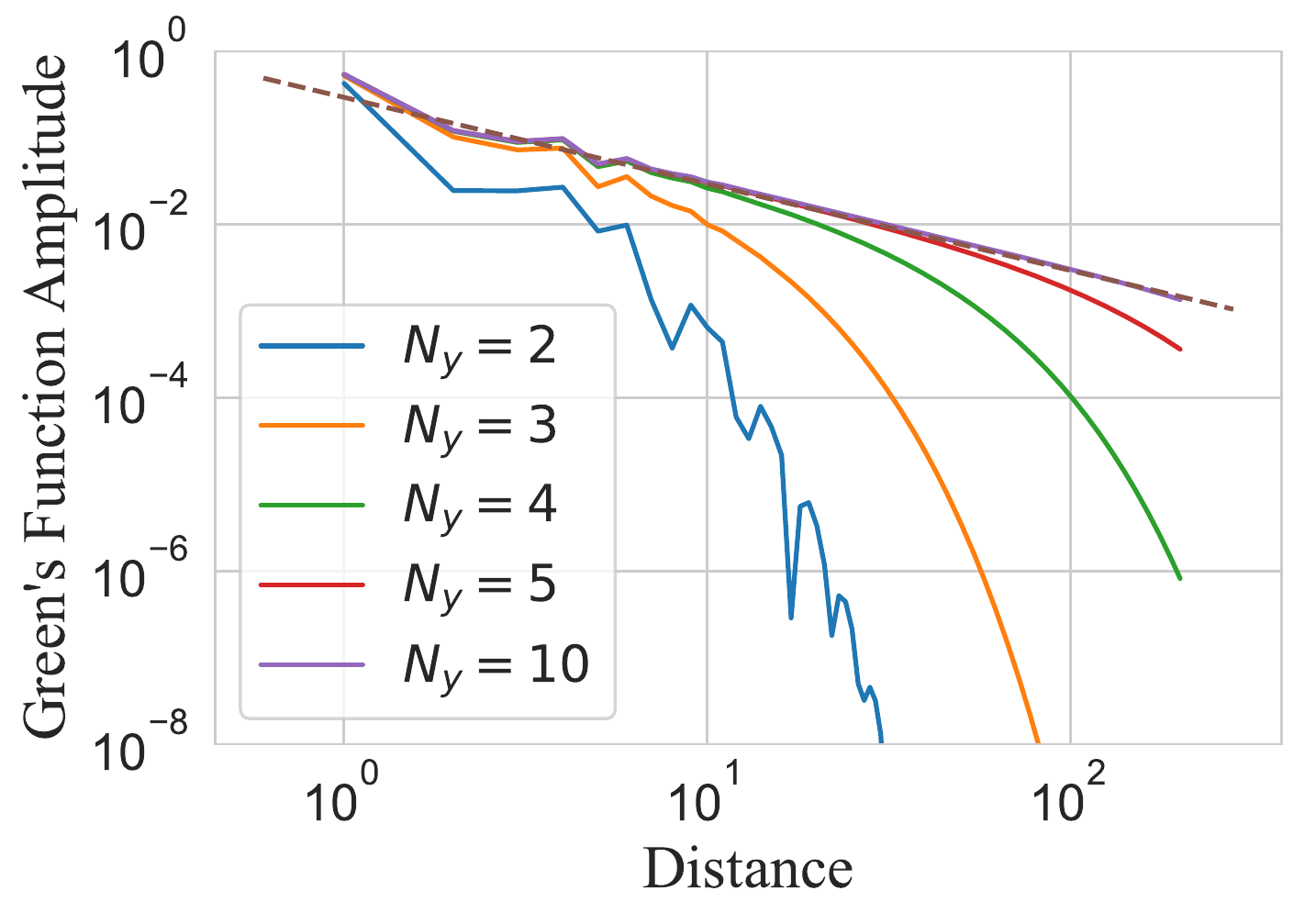}
\par\end{centering}
\caption{The Green's function of the electronic $\beta_{(i,0,A)}$ operator in the SC [Eq.~\eqref{eq:GF}] as a function of distance along the strip for different widths. $1/x$ decay is shown for reference by the dashed line.}
\end{figure}

Next, we show the Green's function of the electronic operator $\beta_{(i,0,A)}$ in the SC as a function of distance along the strip for different widths. For systems with a small width, the Green's function decays exponentially with distance. As the width increases, there is a large crossover regime where the Green's function decays as $1/x$, as expected for a chiral Majorana mode. However, a width of at about $N_y = 5$ or larger is needed to get a clear $1/x$ regime.

\bibliography{References}

\end{document}